\documentstyle[11pt,moriond,epsfig]{article}
\def\beq{\begin{equation}}
\def\eeq{\end{equation}}
\def\bea{\begin{eqnarray}}
\def\eea{\end{eqnarray}}

\begin{document}
\vspace*{4cm}

\vspace*{-4cm}
\noindent
hep-ph/0005199 \\
ANL-HEP-CP-00-057 \\
DESY 00-076 \\[2cm]

\title{PREDICTIONS FOR ASSOCIATED PRODUCTION OF \\ GAUGINOS AND GLUINOS 
AT NLO IN SUSY-QCD}

\author{ E.L. BERGER$^a$, \underline{M. KLASEN}$^b$, AND T. TAIT$^a$ }

\address{$^a$ Argonne National Laboratory, Argonne, IL 60439, USA \\
$^b$ II. Institut f\"ur Theoretische Physik, Universit\"at Hamburg, Luruper
Chaussee 149,\\ D-22609 Hamburg, Germany}

\maketitle\abstracts{NLO SUSY-QCD contributions to associated production of
gluinos and gauginos are shown to enhance the cross sections by about 10\% at
the Tevatron and by as much as a factor of two at the LHC. They shift the mass
determinations or discovery limits, soften the $p_T$ spectra, and stabilize
the predictions against variations of the renormalization and factorization
scales.}

\section{Introduction}
The search for supersymmetry (SUSY) is a major goal of the Tevatron Run
II and LHC physics programs. If SUSY exists at the electroweak scale, SUSY
partners of the Standard Model (SM) particles will either be discovered at
these hadron colliders, or a very large region of SUSY parameter space will be
excluded, provided reliable theoretical predictions in next-to-leading order
(NLO) SUSY-QCD are available \cite{Abel:2000vs,Berger:1999hd,Berger:1999kh,%
Beenakker:1999xh}. We calculate the NLO contributions for associated
production of gauginos and gluinos at hadron colliders \cite{Berger:1999mc,%
Berger:2000xx}. This production channel
is enhanced by the strong coupling of the gluino and by the mass of the
gaugino which is small in many popular models of SUSY breaking. The leptonic
decay of the gaugino makes this process a good candidate for a mass determination 
of the gluino or the discovery or exclusion of a light gluino.

\section{NLO SUSY-QCD Formalism}

The associated production of a gluino and a gaugino proceeds in leading order
(LO) through a quark-antiquark initial state and the
exchange of an intermediate squark in the $t$-channel or $u$-channel.
At NLO, virtual loop corrections must be considered which involve the
exchange of intermediate SM or SUSY particles in self-energy, vertex
correction, or box diagrams. Ultraviolet and infrared divergences appear at
the upper and lower boundaries of integration over unobserved loop momenta.
They are regulated dimensionally and removed through renormalization or
cancellation with corresponding divergences in 2 to 3 parton (real emission)  
diagrams that have an additional gluon radiated into the final state.  At NLO 
there is a second 2 to 3 parton process, one with a $qg$ initial state and 
a light quark emitted into the final state.  In addition to soft divergences, 
real emission contributions have collinear divergences that are factored into
the NLO parton densities.

\section{Tevatron and LHC Cross Sections}

To obtain quantitative predictions for the associated production
of gauginos and gluinos at Run II of the Tevatron and at the LHC, we 
convolve LO and NLO partonic cross sections with 
CTEQ5 parton densities in LO and NLO
($\overline{\rm MS}$)
along with 1- and 2-loop expressions for $\alpha_s$, the corresponding
values of $\Lambda$, and five active quark flavors.  To constrain
the SUSY parameter space, we choose an illustrative SUGRA model with
$m_0=100$ GeV, $A_0$=300 GeV, tan $\beta$ = 4, sgn $\mu$ = +, and we vary
$m_{1/2}$ between 100 and 400 GeV. The resulting masses for
$\tilde{\chi}_{1...4}^0$ vary between 31...162, 63...317, 211...665, and
241...679 GeV, and $\tilde{\chi}_{1,2}^\pm$ are almost mass degenerate with
$\tilde{\chi}^0_{2,4}$. The mass $m_{\tilde{\chi}_3^0}<0$ inside a 
polarization sum.
Our method is not restricted to the SUGRA case and can be applied to any
SUSY breaking model.  

\begin{figure}
\begin{center}
\psfig{figure=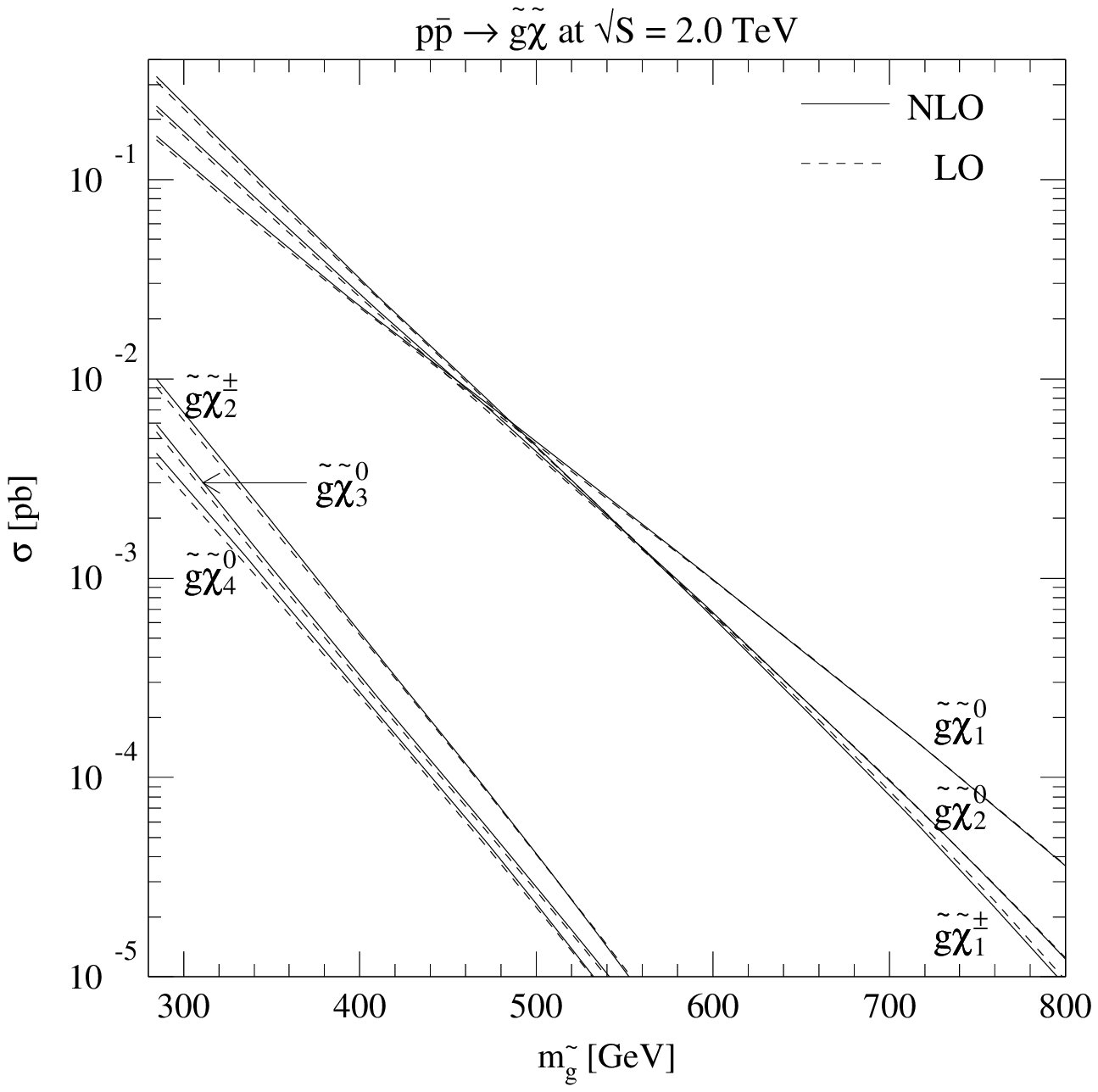,height=7.9cm}
\psfig{figure=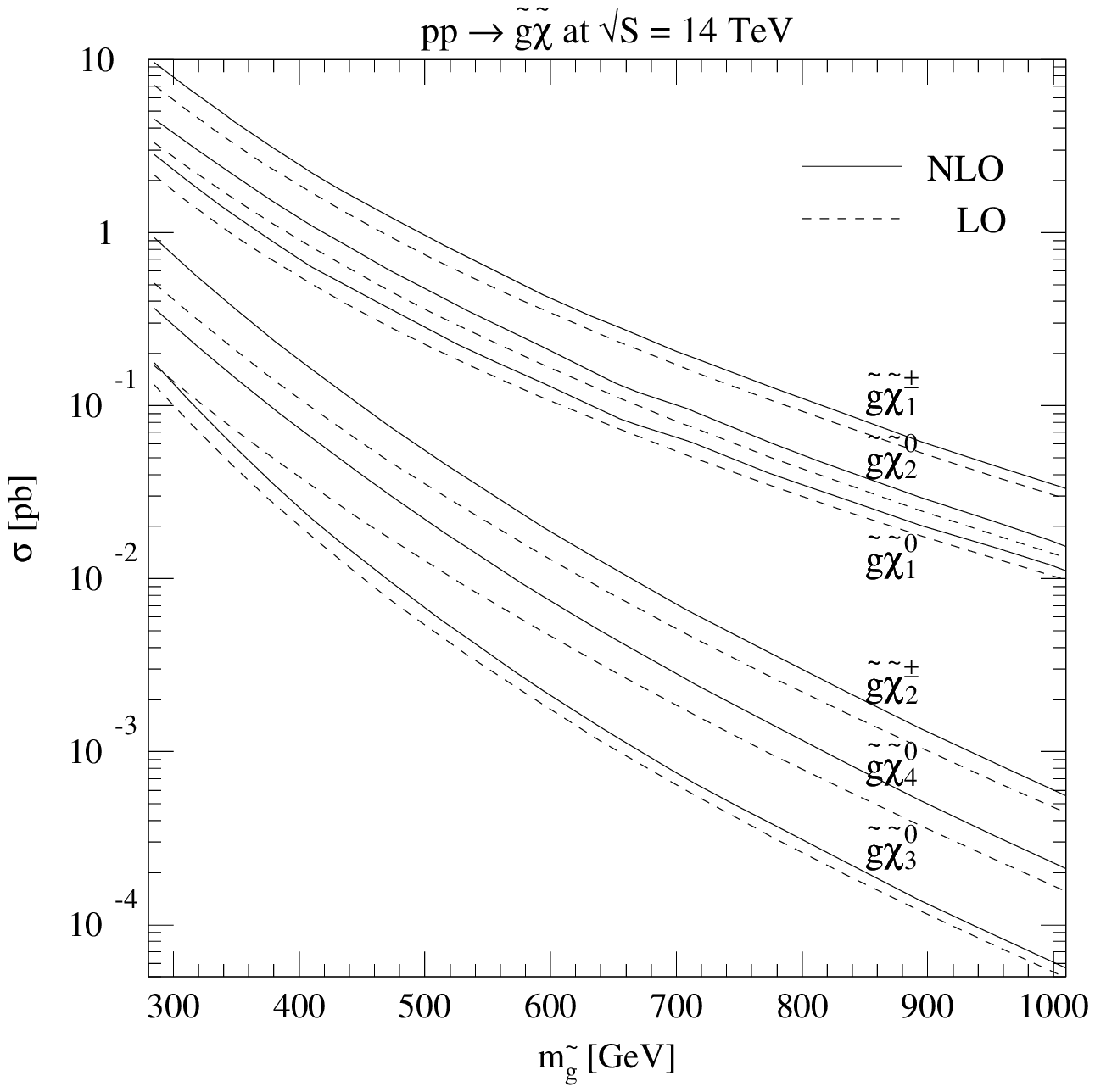,height=7.9cm}
\end{center}
\caption{Total hadronic cross sections at Run II of the Tevatron and at the LHC
in a typical SUGRA model as functions of the gluino mass. 
\label{fig:xsec}}
\end{figure}

We present the total hadronic cross sections in Figure \ref{fig:xsec}
as functions of the gluino mass. The light gaugino channels should be
observable at both colliders, while the heavier Higgsino channels are
suppressed by about one order of magnitude and might be observable only 
at the LHC.

\begin{figure}
\begin{center}
\psfig{figure=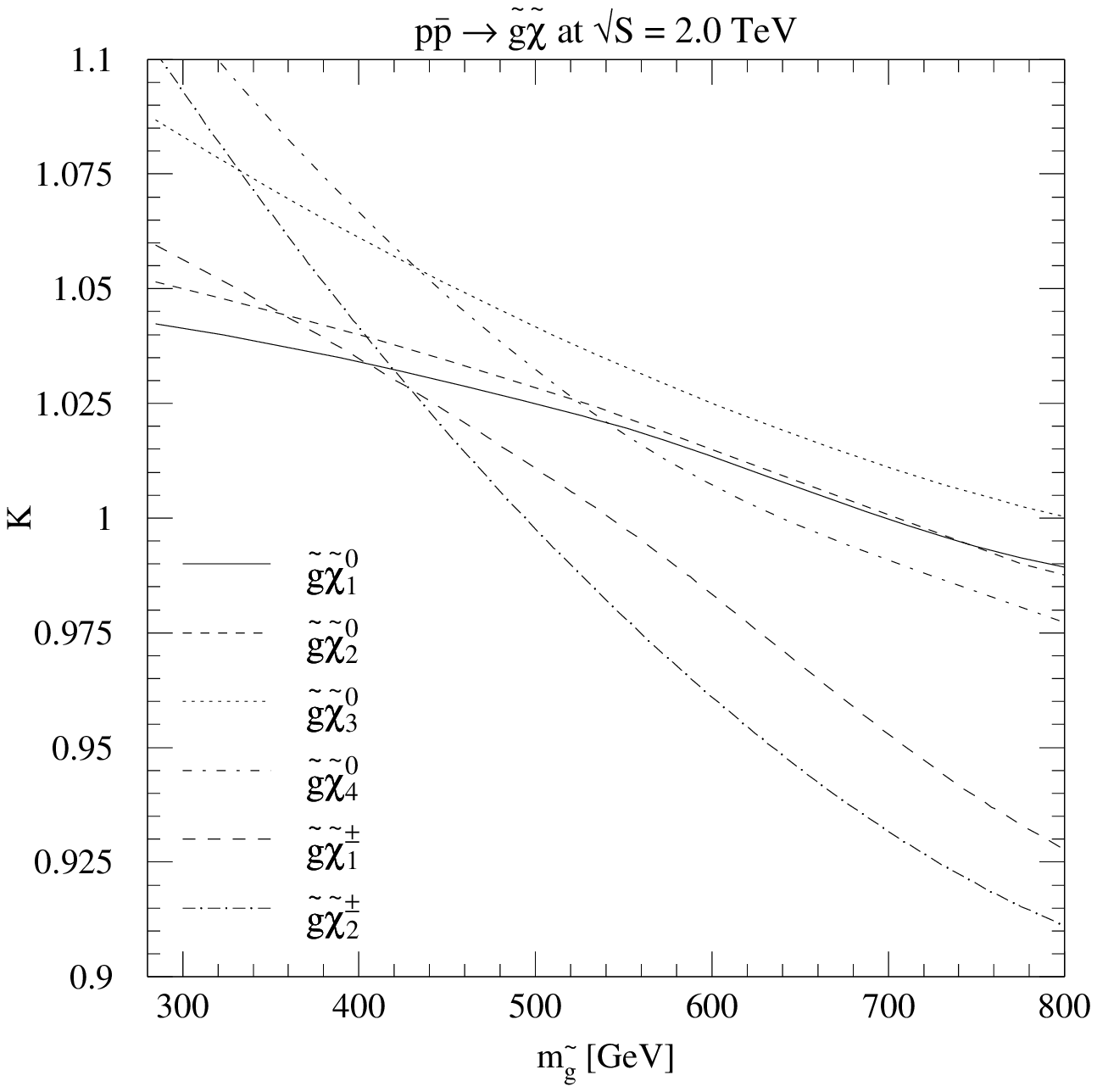,height=7.9cm}
\psfig{figure=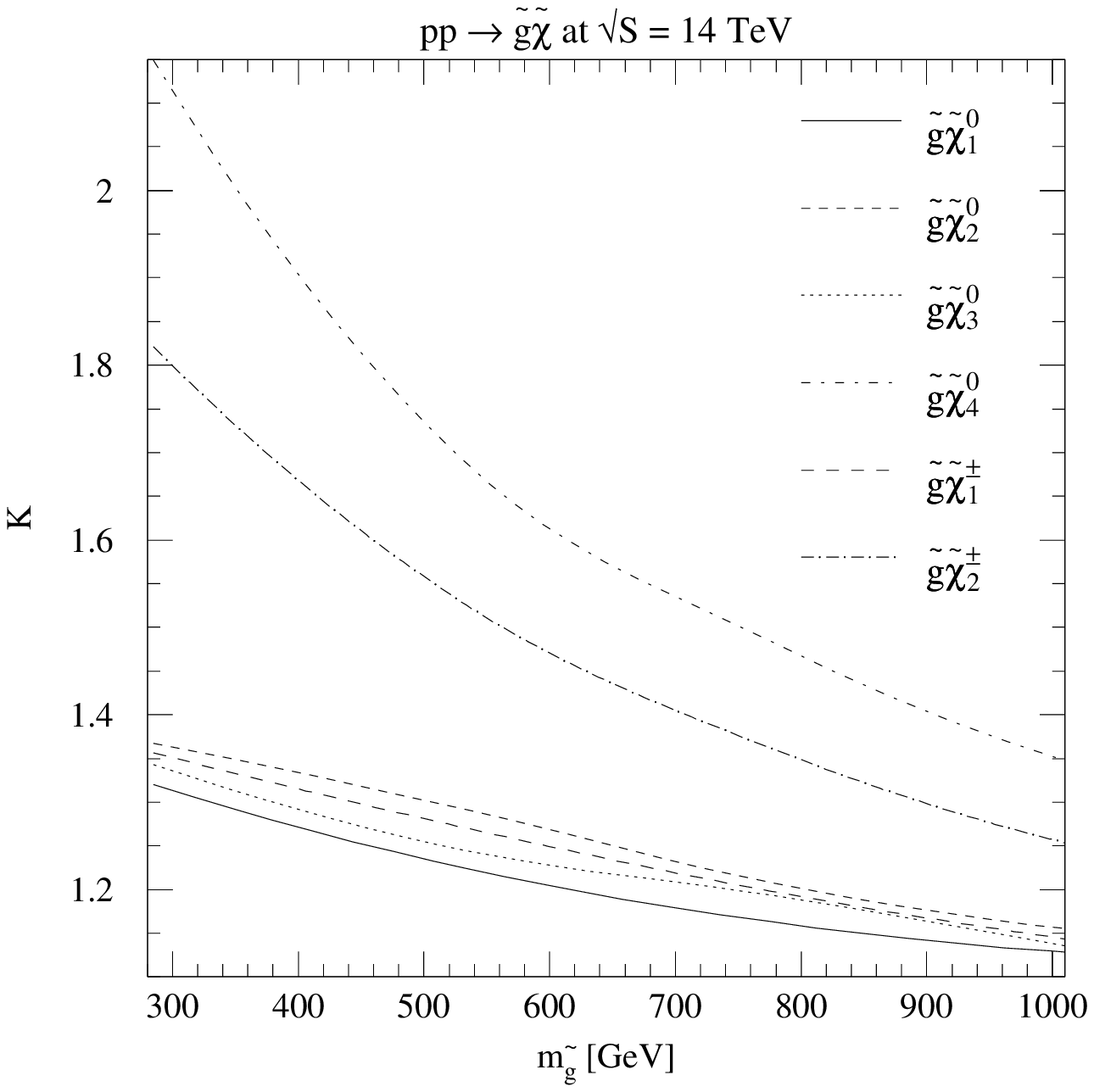,height=7.9cm}
\end{center}
\caption{$K$ factors at Run II of the Tevatron and at the LHC in the SUGRA
model as functions of the gluino mass. 
\label{fig:kfac}}
\end{figure}

The impact of the NLO corrections can be seen more readily in the ratio of NLO
to LO cross sections computed at a renormalizaton scale set equal to the average 
mass of the final state particles. Figure \ref{fig:kfac}
shows that the NLO effects are moderate (of ${\cal O}$ (10\%)) at the Tevatron,
while at the LHC the NLO contributions can increase the cross sections by as
much as a factor of two. Enhancements of this size can shift mass
determinations or discovery limits for SUSY particles by tens of GeV and must
therefore always be taken into account.

\begin{figure}[t]
\begin{center}
\psfig{figure=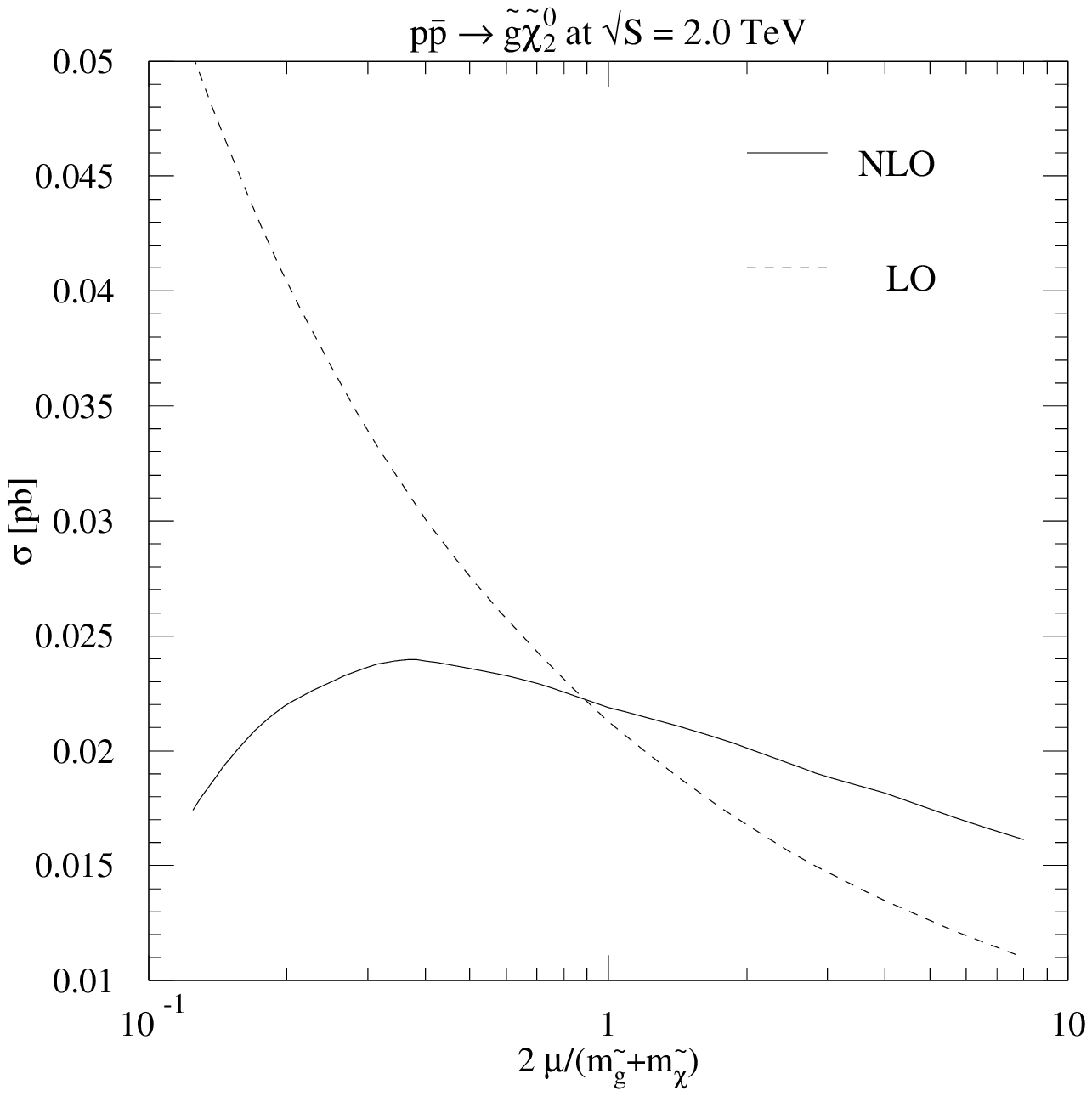,height=7.9cm}
\psfig{figure=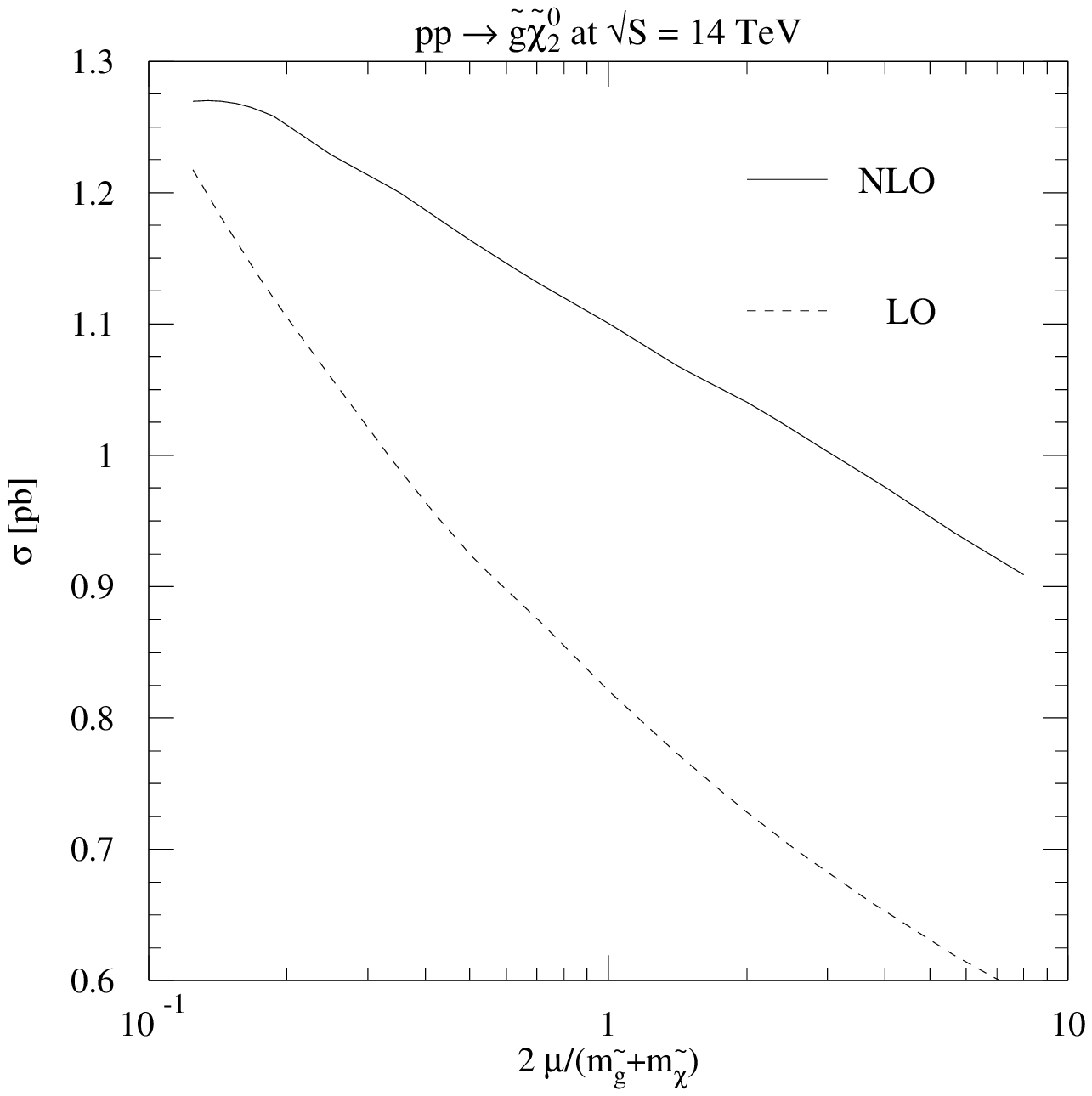,height=7.9cm}
\end{center}
\caption{Dependence of the total $\tilde{g}\tilde{\chi}_2^0$ cross section
in the SUGRA model 
on the common renormalization and factorization scale $\mu$. The LO dependence
is reduced considerably in NLO.
\label{fig:mu}}
\end{figure}

An important measure of the theoretical uncertainty is the variation of the
hadronic cross section with the renormalization and factorization scales.
At LO, these scales enter only in the strong coupling constant
and the parton densities, while
at NLO they appear also explicitly in the hard cross section. As a result,
the scale dependence is reduced considerably, as can be seen in Figure
\ref{fig:mu}.
The Tevatron (LHC) cross sections vary by $\pm 23 (12) \%$ at LO, but only by
$\pm 8 (4.5) \%$ in NLO when the scale is varied by a factor of two around
the central scale.

\begin{figure}
\begin{center}
\vspace*{-2.5cm}
\psfig{figure=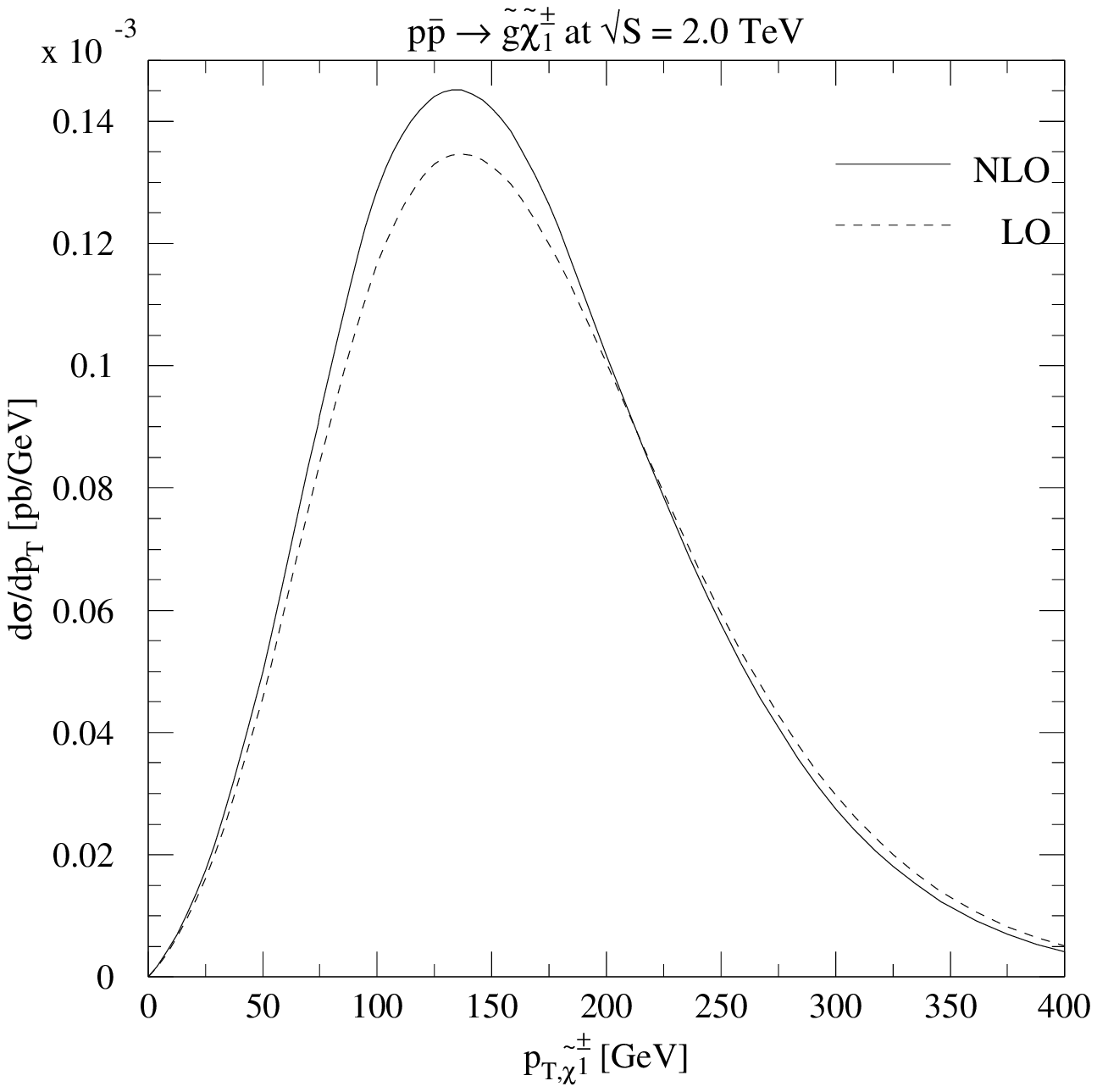,height=7.9cm}
\psfig{figure=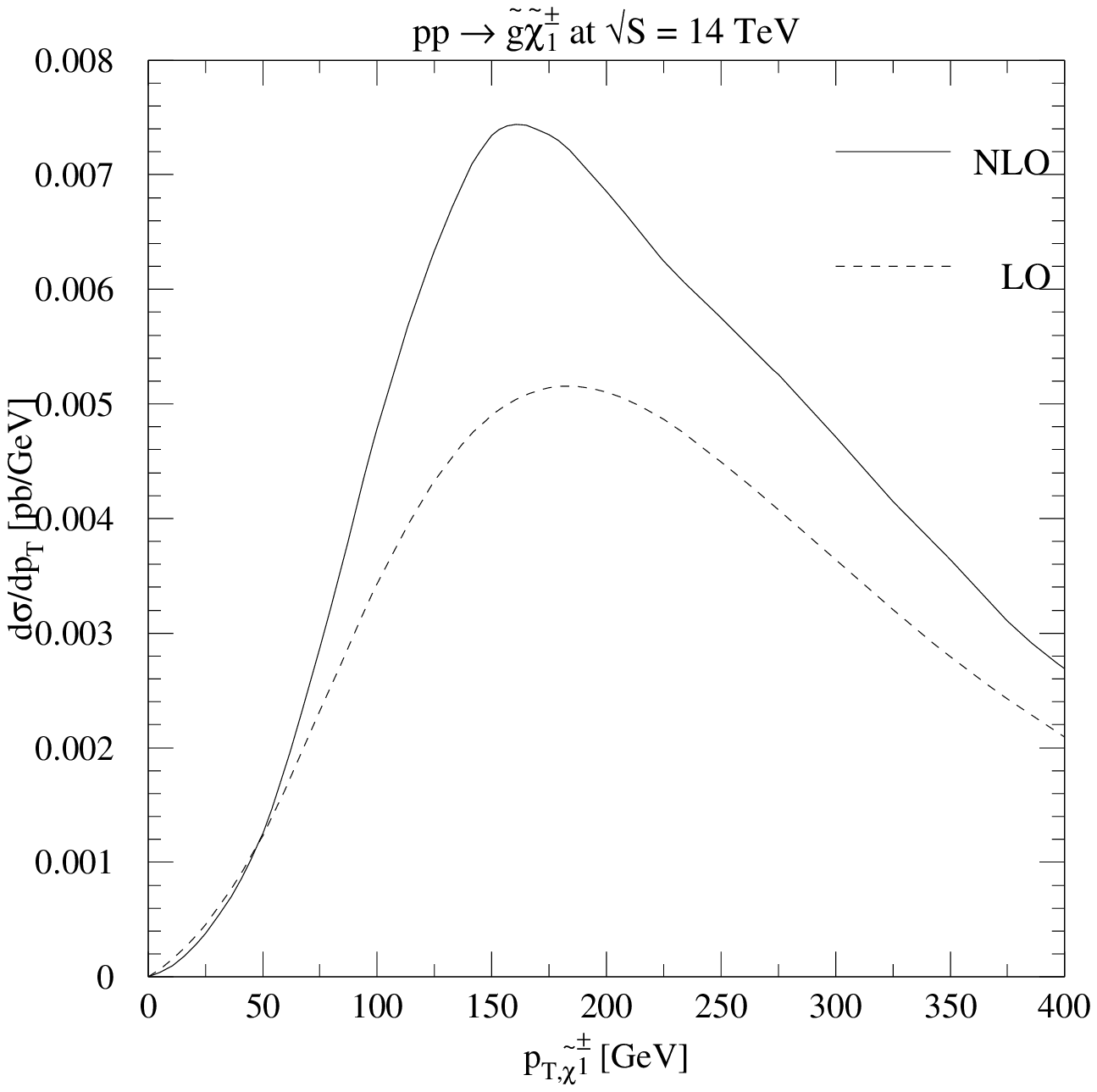,height=7.9cm}
\end{center}
\caption{Dependence of the cross section for $\tilde{g}\tilde{\chi}_1^\pm$
in the SUGRA model on the transverse momentum of the light
chargino. The NLO distributions are shifted
to lower $p_T$ with respect to LO. At the LHC, the NLO enhancement is more
prominent.
\label{fig:pt}}
\end{figure}

For experimental searches, distributions in transverse momentum are 
important since cuts on $p_T$ help to enhance the signal. In Figure
\ref{fig:pt}
we demonstrate that NLO contributions can have a large
impact on $p_T$ spectra, especially at the LHC, where contributions from
the $gq$ initial state become important.  At the Tevatron the
NLO $p_T$-distribution is shifted to lower $p_T$ with respect to the LO
expectation.

\section{Conclusions}

The direct search for SUSY particles is a major 
goal at hadron colliders. For associated production 
of gauginos and gluinos, we demonstrate that NLO SUSY-QCD
corrections stabilize the LO estimates against variations of the
renormalization and factorization scale.  The cross sections
are increased by as much as a factor of two at the LHC, and the $p_T$
spectra are softened.

\section*{Acknowledgments}
M.K.\ thanks the organizers of the XXXVth Rencontres de Moriond for the kind
invitation and financial support. Work at ANL is supported by the U.S.\
Department of Energy, Division of High Energy Physics,
under Contract
W-31-109-ENG-38. M.K.\ is supported by the Bundesministerium f\"ur Bildung
und Forschung under Contract 05 HT9GUA 3, by Deutsche Forschungsgemeinschaft
under Contract KL 1266/1-1, and by the European Commission under Contract
ERBFMRXCT980194.

\end{document}